\documentclass[journal=jacsat,manuscript=article]{achemso}

\usepackage[version=3]{mhchem} 
\usepackage[normalem]{ulem}

\usepackage{xcolor}

\author{Marcin Rosmus}
\email{marcin.rosmus@universite-paris-saclay.fr}
\affiliation{Universit\'{e} Paris-Saclay, CNRS, Institut des Sciences Mol\'{e}culaires d’Orsay, 91405, Orsay, France}
\alsoaffiliation{Solaris National Synchrotron Radiation Centre, Jagiellonian University, Czerwone Maki 98, 30-392 Krak\'{o}w, Poland}
\alsoaffiliation{Marian Smoluchowski Institute of Physics, Jagiellonian University, Prof. S. {\L}ojasiewicza 11, PL-30348 Krak\'{o}w, Poland}

\author{Natalia Olszowska}
\affiliation{Solaris National Synchrotron Radiation Centre, Jagiellonian University, Czerwone Maki 98, 30-392 Krak\'{o}w, Poland}

\author{Rafa\l{} Kurleto}
\affiliation{Marian Smoluchowski Institute of Physics, Jagiellonian University, Prof. S. {\L}ojasiewicza 11, PL-30348 Krak\'{o}w, Poland}
\alsoaffiliation{Solaris National Synchrotron Radiation Centre, Jagiellonian University, Czerwone Maki 98, 30-392 Krak\'{o}w, Poland}

\author{Zbigniew Bukowski}
\affiliation{Insitute of Low Temperature and Structure Research, Polish Academy of Sciences, P.O. Box 1410,50-950 Wroc\l{}aw, Poland}

\author{Pawe\l{} Starowicz}
\email{pawel.starowicz@uj.edu.pl}
\affiliation{Marian Smoluchowski Institute of Physics, Jagiellonian University, Prof. S. {\L}ojasiewicza 11, PL-30348 Krak\'{o}w, Poland}

\title{Observation of the Dirac Dispersions in Co-doped CaFe$_{2}$As$_{2}$}

\abbreviations{ARPES,SDW,PM, AFM, BZ, FS}

\begin{document}


\begin{abstract}
We performed an angle-resolved photoemission spectroscopy (ARPES) study of the electronic structure of the CaFe$_{2}$As$_{2}$ 122-iron pnictide, a parent compound, and two iron-based superconductors CaFe$_{2-x}$Co$_{x}$As$_{2}$ (x= 0.07 and 0.15). We studied the band structure of this system across the phase diagram with the transition from the orthorhombic spin density wave (SDW) phase to the tetragonal paramagnetic phase. We observed characteristic features of the electronic structures corresponding to the antiferromagnetic phase in the parent compound and the samples with low cobalt concentration ($x = 0.07$). For highly doped systems  ($x = 0.15$), the measurements revealed  the concentric branches of the Fermi surface, which are associated with paramagnetic and superconducting 122-iron pnictides. We found  the existence of Dirac cones located at 30 meV below Fermi energy for nonsuperconducting CaFe$_{2}$As$_{2}$ and superconducting CaFe$_{1.93}$Co$_{0.07}$As$_{2}$ orthorhombic SDW systems.
\end{abstract}

\section{Introduction}

The discovery of superconductivity in iron-based compounds opened a new chapter in the research of strongly correlated matter. One of the most interesting families among these materials is so-called 122 family with the structure type of BaFe$_{2}$As$_{2}$. It exhibits the structural transition between the tetragonal paramagnetic phase (PM) and the orthorhombic spin density wave (SDW) phase, which is often named antiferromagnetic (AFM) \cite{1ni_tuning_2011,2tai_calculated_2013,3chu_determination_2009, PhysRevB.89.020511}. Its interesting phase diagram together with the possibility of unconventional pairing mechanism made this system a compelling object of investigation. For a higher concentration of impurities (or a higher external pressure~\cite{4widom_first-principles_2013}), the SDW phase disappears, and the system remains in the tetragonal phase for the entire temperature range; however, a significant shortening of the lattice constant c is observed \cite{PhysRevB.93.024516, PhysRevLett.103.026404, doi:10.1143/JPSJ.80.103701}, which is often called the transition to the collapsed tetragonal phase. Interestingly, the superconductivity is observed in a wide range of phase diagram, extending over both the SDW and PM regions. 

Experimental works~\cite{5richard_observation_2010, 6matusiak_nernst_2010,7xu_optical_2018,8terashima_fermi_2018,9chen_two-dimensional_2017,10matusiak_-plane_2019,11huynh_both_2011,12sutherland_evidence_2011} report the existence of nontrivial topological states in the family of 122 compounds. Theoretical predictions suggest the existence of Dirac cones with the same chirality~\cite{13ran_nodal_2009}, furthermore theoretical calculations for this family of compounds~\cite{13ran_nodal_2009,14pan_evolution_2013} show that the occurrence of Dirac states in this family is related to the physical symmetry and topology of the band structure, which stabilizes the ground state of the gapless spin density waves (SDWs) phase. To this day Dirac states in the CaFe$_{2}$As$_{2}$ compound have not been reported. 

\begin{figure}[b]
\includegraphics[width=\linewidth]{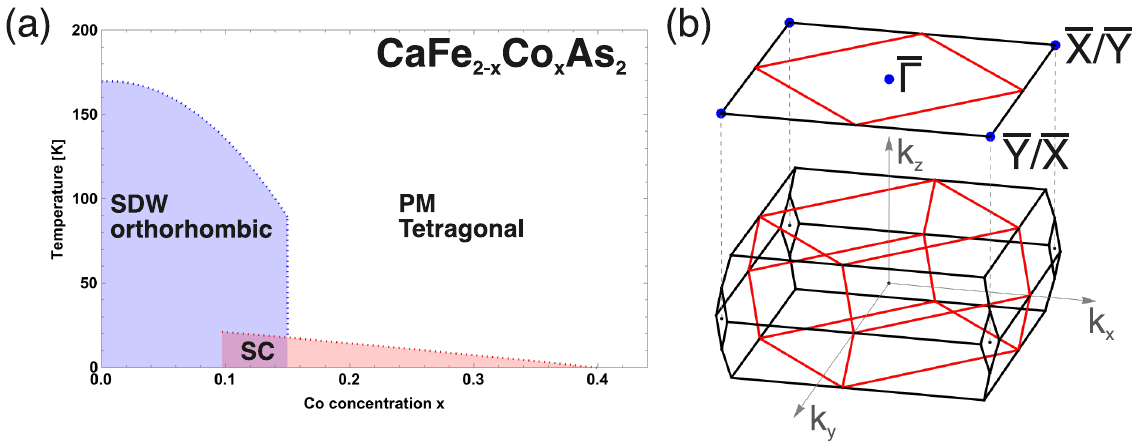}%
\caption{(a) Schematic phase diagram of the cobalt doped CaFe$_{2}$As$_{2}$ system. (b) Brillouin zone of CaFe$_{2}$As$_{2}$ in tetragonal (black) and orthorhombic (red) phase.
}
\label{f1}
\end{figure}

\begin{figure*}[t]
\includegraphics[width=12cm]{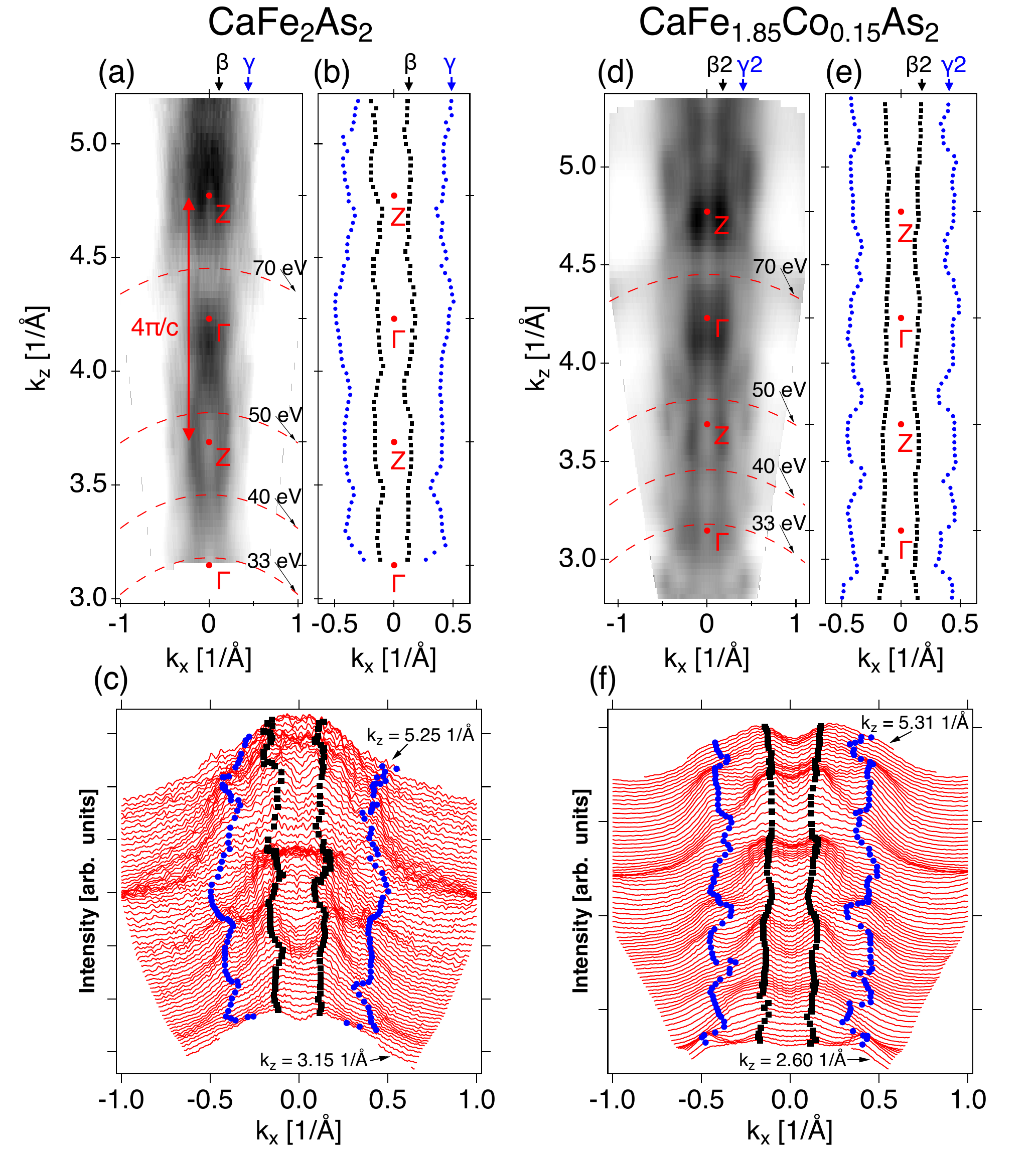}%
\caption{FS map along k$_{z}$ - k$_{x}$ plane for CaFe$_{2}$As$_{2}$ and CaFe$_{1.85}$Co$_{0.15}$As$_{2}$ obtained from measurements along k$_{x}$ as defined in Figure~\ref{f1} performed for different photon energies. Selected photon energies correspond to red dashed lines. (a)~Fermi Surface map for CaFe$_{2}$As$_{2}$, (b) the positions of $\beta$ and $\gamma$ bands at the Fermi Surface determined from  MDC curves, (c) MDC waterfall plot for CaFe$_{2}$As$_{2}$. (d) Fermi Surface map for CaFe$_{1.85}$Co$_{0.15}$As$_{2}$, (e)~the positions of $\beta$ and $\gamma$ bands at the Fermi Surface determined from MDC curves, (f) MDC waterfall plot for CeFe$_{1.85}$Co$_{0.15}$As$_{2}$. The measurement for CaFe\(_2\)As\(_2\) was performed at a temperature of 8~K with horizontal linear polarization, for CaFe\(_{1.85}\)Co\(_{0.15}\)As\(_2\), it was conducted at 15~K, also using horizontal linear polarization.
}
\label{f2}
\end{figure*}

The aim of this work was to study the electronic structure of the cobalt doped CaFe$_{2}$As$_{2}$ compound in both the antiferromagnetic and the paramagnetic regions of the phase diagram. The measurements were performed for the samples with three different compositions: parent compound CaFe$_{2}$As$_{2}$, lightly doped CaFe$_{1.93}$Co$_{0.07}$As$_{2}$ and heavily doped CaFe$_{1.85}$Co$_{0.15}$As$_{2}$. The lightly doped sample corresponds to the Co concentration for which the superconducting phase appears in the SDW state, while the heavily doped sample is in the paramagnetic superconducting phase (Figure~\ref{f1}a). The parent compound crystallizes in the tetragonal I4/mmm space group (No. 139) with lattice parameters $a = 3.89245(8)$~\AA~and $c = 11.6403(7)$~\AA~\cite{15saparov_complex_2014, 16ni_first-order_2008}. In the cooling process, there is a structural transition to the orthorhombic phase (Fmmm space group) at approximately 170~K~\cite{17ali_emergent_2017,18tompsett_dimensionality_2010,19watson_probing_2019, 23wang_symmetry-broken_2013}. Highly doped samples do not show this transition, they remain in the tetragonal phase. The first Brillouin zones (BZ) for the tetragonal and the orthorhombic phases are shown in Figure~\ref{f1}b. 

In this article, systematic angle-resolved photoemission spectroscopy (ARPES) studies revealed band structures and Fermi surfaces for tetragonal and orthorhombic systems. Dirac cones were found in the orthorhombic samples.

\begin{figure*}[t]
\includegraphics[width=\textwidth]{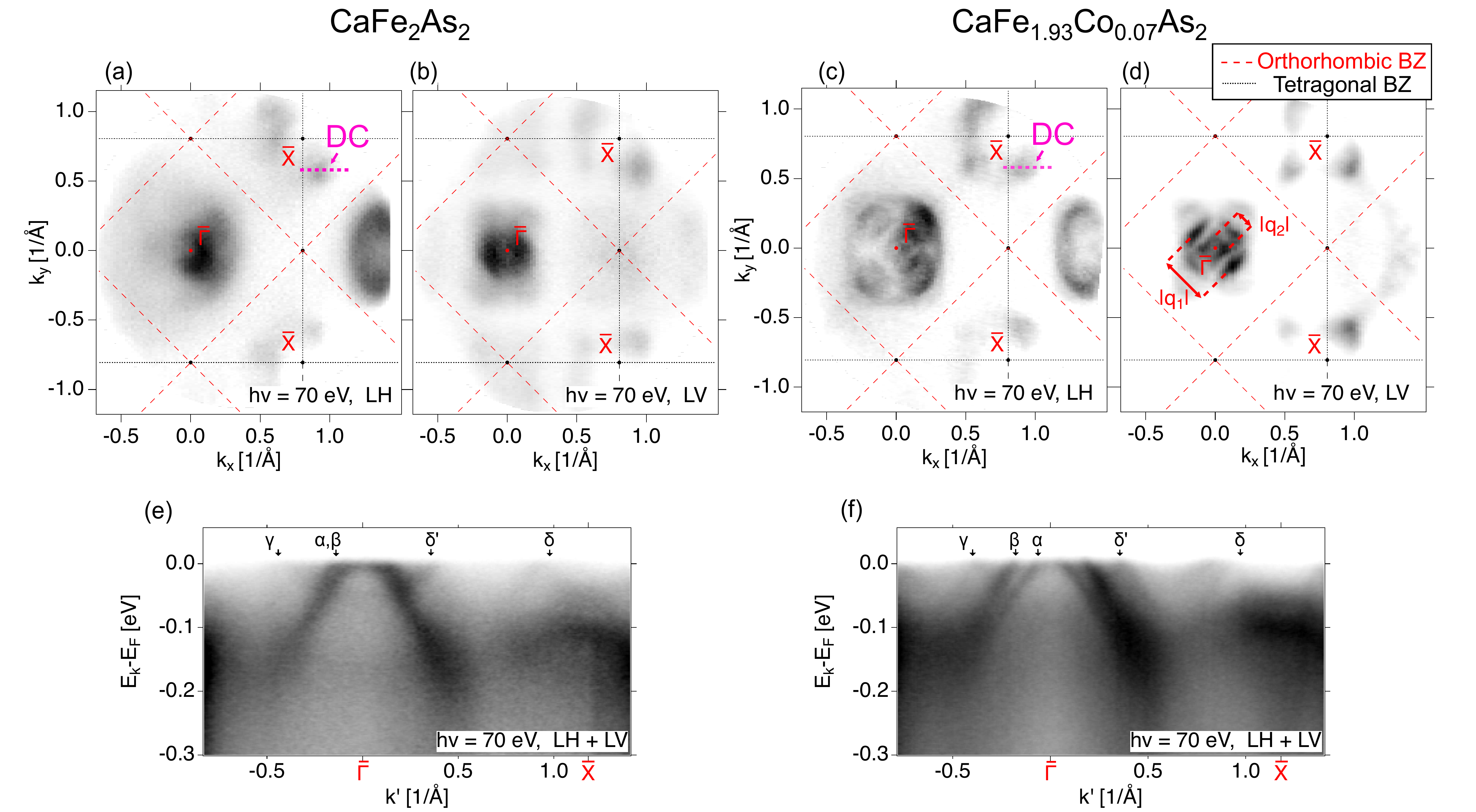}%
\caption{Fermi surfaces and band structure obtained with photon energy of 70 eV and different polarizations for CaFe$_{2}$As$_{2}$ (left) and CaFe$_{1.93}$Co$_{0.07}$As$_{2}$ (right). The Brillouin zones for tetragonal and orthorhombic phases are drawn with black dotted and red dashed lines, respectively. Details concerning energy and polarization are given in the figures. The paths, for which Dirac cones (DC) have been observed are marked with pink dashed lines in (a) and (c). Nesting vectors $q_1$ and $q_2$ are indicated in (d). 
(e) and (f) show band structure along $\Gamma - \bar{\text{X}}$ obtained by adding the data for linear vertical (LV) and linear horizontal (LH) polarizations. The observed $\alpha$, $\beta$, $\gamma$, $\delta$ and $\delta'$ bands are marked.}
\label{f3}
\end{figure*}

\section{Methods and techniques}
\label{sec.method}
Single crystals of CaFe$_{2-x}$Co$_{x}$As$_{2}$ were grown using Sn flux method. The components were loaded into alumina crucibles and placed in quartz ampoules that were evacuated and filled with Ar gas under a pressure of 0.3~bar and sealed. The ampoules were heated to $600^{\circ}$C for 4~h, kept there for 1~h, then heated in 5~h to $1050^{\circ}$C, and kept at that temperature for 5~h so that all components are dissolved in the Sn flux. Next, the ampoules were cooled slowly at 2~K/h down to $600^{\circ}$C and subsequently cooled to room temperature. The flux was removed by a treatment with dilute hydrochloric acid. Phase purity was checked by powder X-ray diffraction (XRD), and the chemical composition of the doped crystals was determined by energy dispersive XRD analysis. The transport properties of CaFe$_{2-x}$Co$_{x}$As$_{2}$ single crystals have been characterized by measuring the resistivity and AC susceptibility as a function of temperature. For the heavily doped sample CaFe$_{1.85}$Co$_{0.15}$As$_{2}$, the transition to SC is observed at the temperature 
$T_{C,onset} = 18.7$~K, while for the lightly doped sample CaFe$_{1.93}$Co$_{0.07}$As$_{2}$ at the onset temperature $T_{C,onset} = 18$~K, and no flattening of resistance is observed.  However, for a lightly doped sample, resistance measurements revealed a transition to the SDW state at $T_{SDW} = 135 - 140$~K.
ARPES measurements were performed at URANOS beamline in the Solaris synchrotron~\cite{Szlachetko2023} in Kraków, Poland. The endstation was equipped with the Scienta Omicron DA30L photoelectron spectrometer and the radiation source was a quasi-periodic APPLE II undulator. The samples were cleaved in situ under ultra-high vacuum at room temperature and the measurements were made at the temperature of 12~K. Base pressure was below $5 \times 10^{-11}$~mbar. The estimated resolution of the beamline varied from 7~meV at a photon energy of 33~eV to 20~meV at 70~eV. Both linear vertical (LV) and linear horizontal (LH) polarizations were used. The desired orientation of single crystals was obtained using the Laue diffraction method.

\begin{figure*}[t]
\includegraphics[scale=0.22]{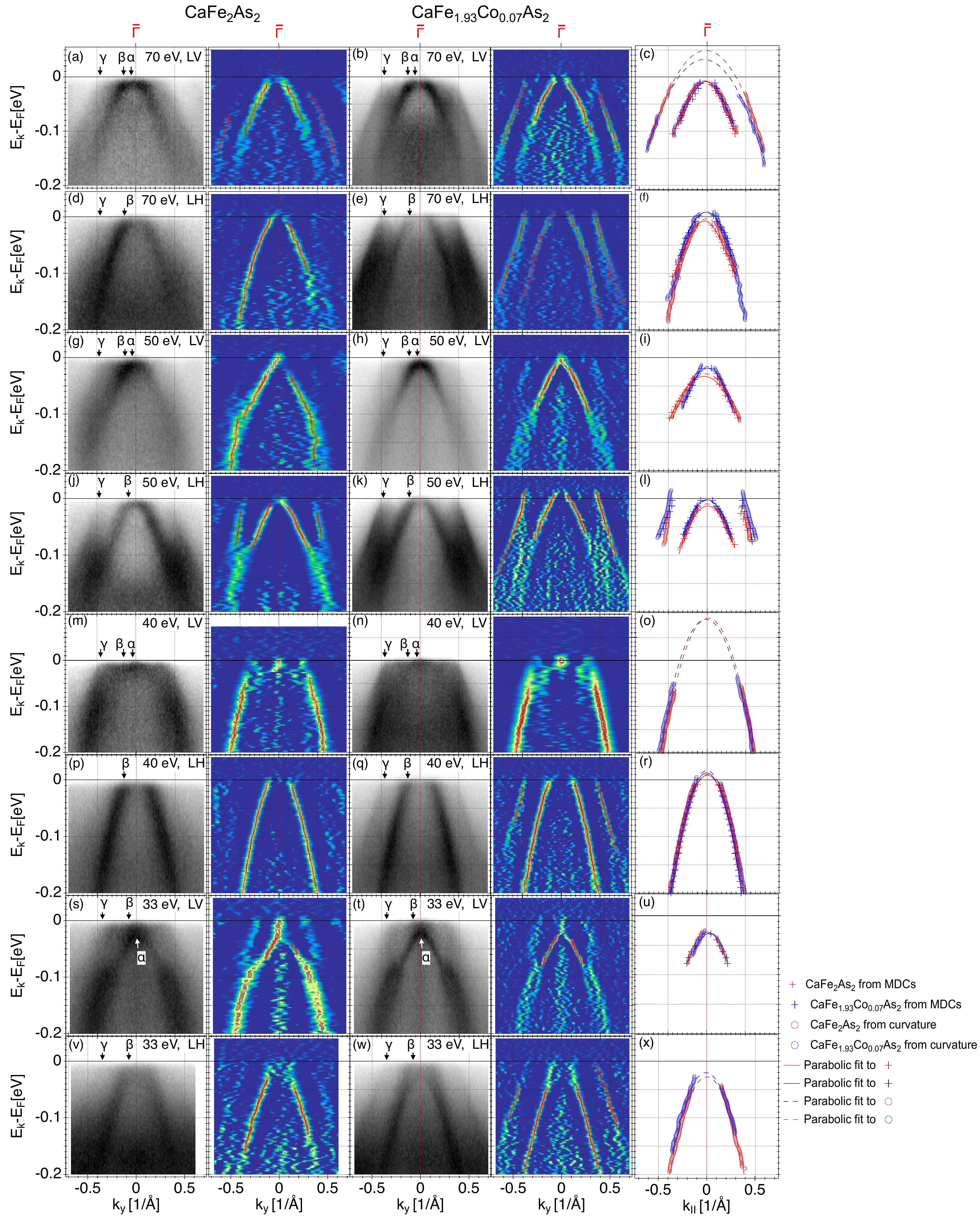}%
\caption{ARPES spectra and their curvatures collected at several photon energies and for LV and LH polarizations for CaFe$_{2}$As$_{2}$ (left) and CaFe$_{1.93}$Co$_{0.07}$As$_{2}$ (right). The last column shows a comparison of band positions based on fits to the raw data MDC curves and curvature MDCs. Details regarding energy and polarization are provided in the figures.}
\label{f4}
\end{figure*}

\begin{figure*}
\includegraphics[width=\textwidth]{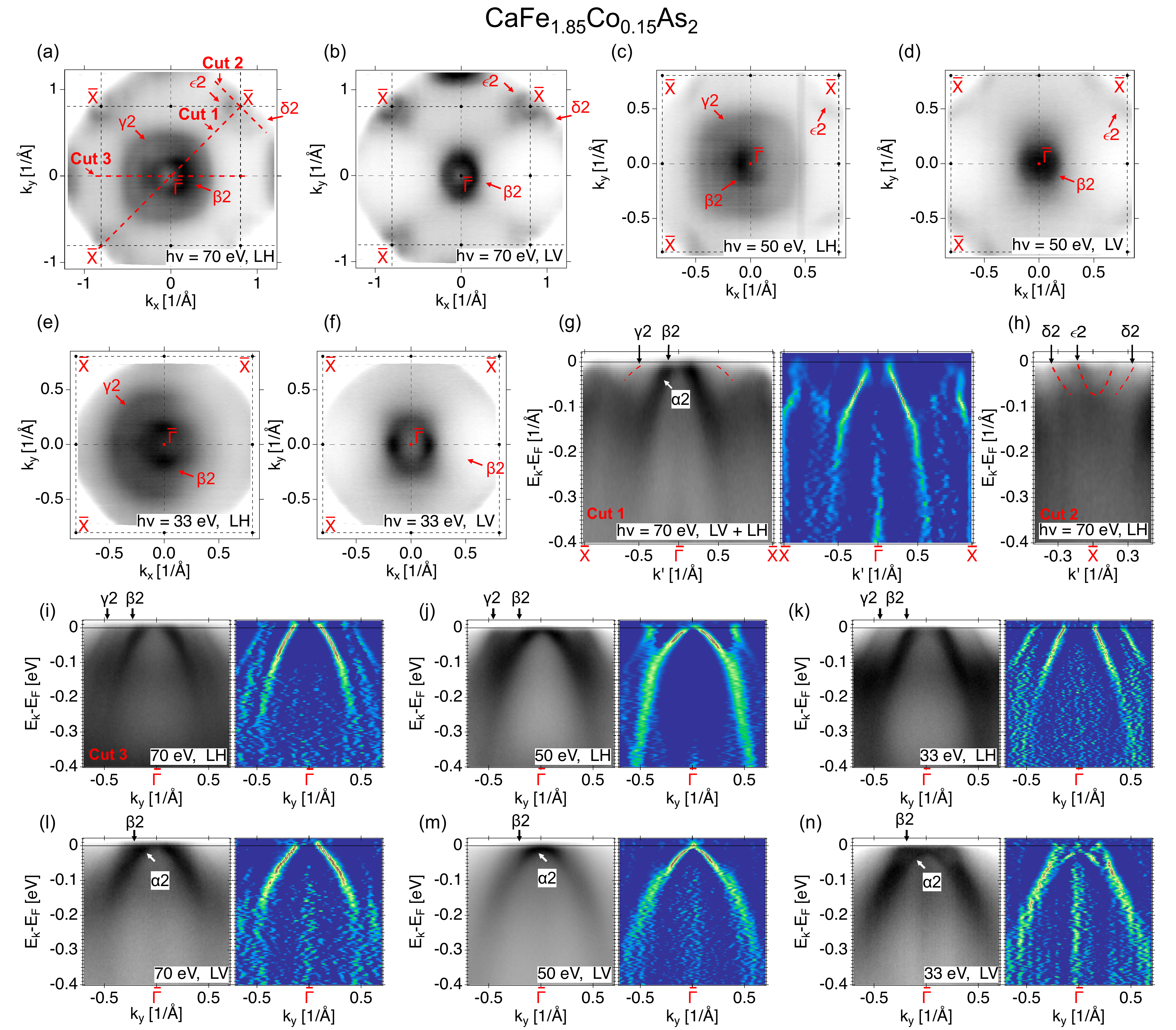}%
\caption{ARPES measurements for the heavily doped sample CaFe$_{1.85}$Co$_{0.15}$As$_{2}$. (a-f) - Fermi surface maps collected using different energies and for different polarizations. Spectra obtained along the high symmetry directions and the corresponding curvatures are shown for $\bar{\Gamma} - \bar{\text{X}}$ (g,h) and along $k_y$ (i-n) -  direction, where $k_y$ is defined in Figure~\ref{f1}. Raw data (gray) are accompanied by curvatures (colour). The band structure along the paths named in (a) as cut 1, cut 2 and cut 3 is presented in (g), (h) and (i), respectively. $\alpha2$, $\beta2$, $\gamma2$, $\delta2$ and $\epsilon2$ bands for a tetragonal system are marked at the dispersions. Photon energies and polarizations are given in the graphs.}
\label{f5}
\end{figure*}

\section{Results and discussion}
\label{sec.res}
Fermi Surface (FS) maps of undoped CaFe$_{2}$As$_{2}$ in the orthorhombic SDW phase and heavily doped CaFe$_{1.85}$Co$_{0.15}$As$_{2}$ in the tetragonal paramagnetic phase, obtained in the $\text{k}_{z}-\text{k}_{x}$ plane (${k}_{z}$ is perpendicular to the surface) are shown in Figure~\ref{f2}. The observed structure is dominated by two bands, named $\beta$ and $\gamma$. The innermost $\alpha$ band is often below the Fermi energy ($E_{F}$) as discussed further in the text. FS branches related to $\beta$ and $\gamma$ extend along $\text{k}_{z}$ direction. To fully unravel the band structure in this direction, the four Gaussians were fitted together with the polynomial background to each momentum distribution curve (MDC). The positions of the maxima obtained in this way are shown in Figure~\ref{f2} with black ($\beta$ pocket) and blue ($\gamma$ pocket) squares. For both samples, the obtained structure is similar, with practically two-dimensional $\beta$ pocket and a periodic $\gamma$  pocket with the period equal to $4\pi/c$ corresponding to the distance between $\Gamma$ points in neighboring BZs along $k_z$ direction. Based on the obtained data, it was possible to find the location of the high symmetry points. For $\Gamma$ point the attributed k vectors are $k_z  = 3.15$~\AA$^{-1}$ and $k_z  = 4.23$~\AA$^{-1}$ and they correspond to the photon energies $h\nu = 32$~eV and $h\nu=68$~eV, respectively. Similarly, the position of the Z points equals $k_{z}  = 3.69$~\AA$^{-1}$ and $k_z  = 4.77$ \AA$^{-1}$ and the photon energies: $h\nu = 44$~eV and $h\nu = 80$~eV, respectively. All obtained FSs are in line with previous research on this topic for undoped BaFe$_{2}$As$_{2}$ and CaFe$_{2}$As$_{2}$ samples~\cite{21liu_three-_2009,24kondo_unexpected_2010}. It is worth mentioning that there are no significant qualitative differences between the maps for both measured systems. Surprisingly, a significantly better quality of the spectra was obtained for the doped samples. 

\begin{figure*}
\includegraphics[width=\columnwidth]{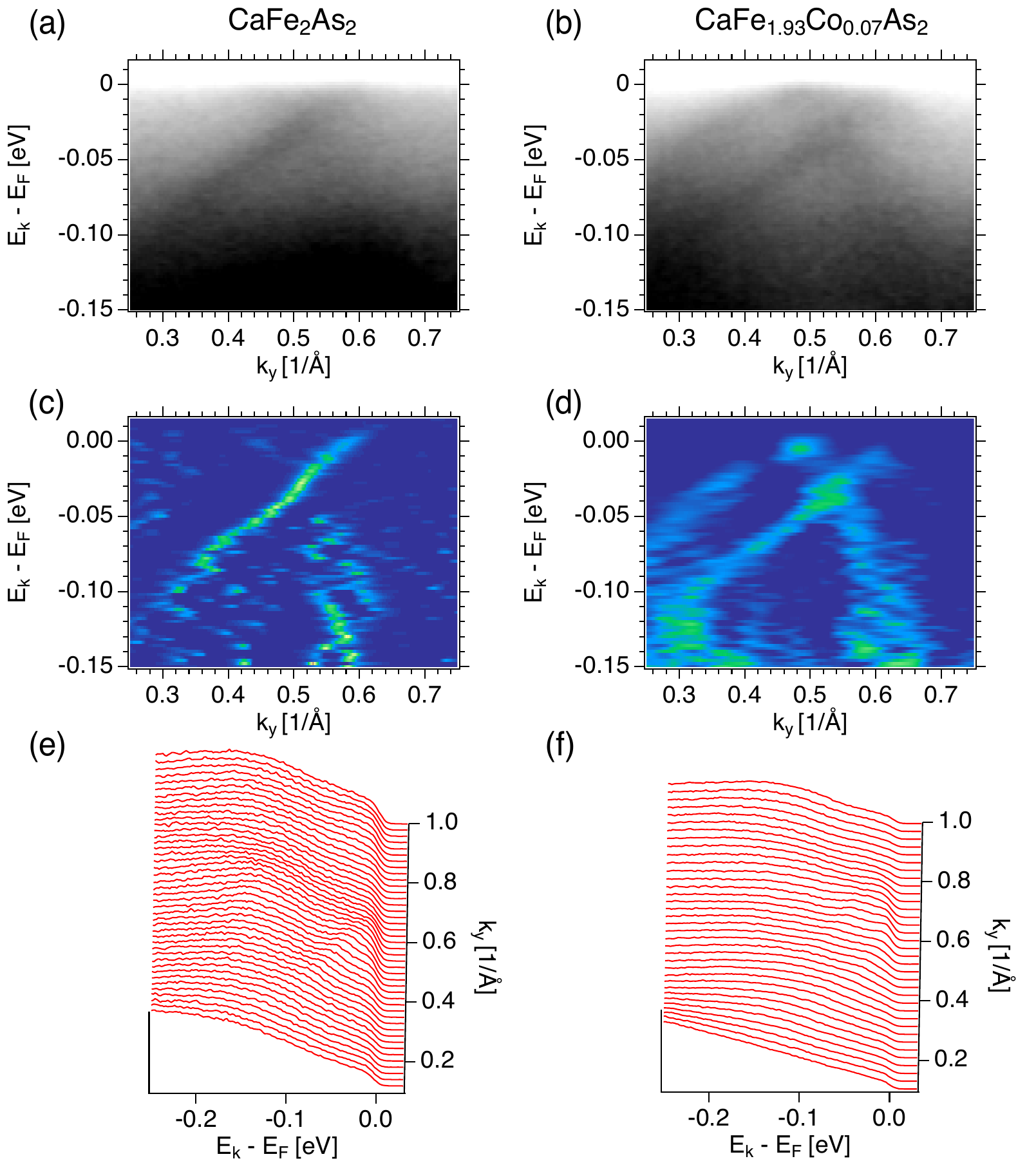}%
\caption{Dirac cones observed near the $\bar{\text{X}}$ point. ARPES spectra of CaFe$_{2}$As$_{2}$ and CaFe$_{1.93}$Co$_{0.07}$As$_{2}$ are shown in (a) and (b), the corresponding curvatures in (c) and (d), and energy distribution curves in (e) and (f), respectively. Measurements were performed at 70 eV with LH  polarization.}
\label{f6}
\end{figure*}

Fermi surface maps along $\text{k}_{x}-\text{k}_{y}$ plane 
of the parent compound CaFe$_{2}$As$_{2}$ and the lightly doped sample CaFe$_{1.93}$Co$_{0.07}$As$_{2}$ are shown for h$\nu$=70 eV photon energy in Figure~\ref{f3} together with  band structure along $\bar\Gamma - \bar{\text{X}}$ direction. Two BZs are marked in Figure~\ref{f3}, the first one represents the tetragonal paramagnetic and superconducting phase (black dashed lines), and the second one is connected to the orthorhombic SDW phase (red dashed lines). The effect of breaking the rotational symmetry, reported earlier~\cite{25yi_symmetry-breaking_2011}, would be obtained if the experiment was performed on a detwinned sample or if domains were significantly larger than the size of the beam. In the FS maps there are no discernible differences between the undoped and lightly doped sample. The bands crossing the Fermi energy (E$_{F}$) around the {$\bar\Gamma$} point create a characteristic petal-like shape, while around the $\bar{\text{X}}$ point, a four-dot structure can be observed. The petal-like shape is associated with the occurrence of the SDW phase~\cite{24kondo_unexpected_2010,22chen_electronic_2011,23wang_symmetry-broken_2013}, and the large part of the FS exhibits nesting. The length of the nesting vectors has been estimated to be $q_{1} \sim 0.31 \cdot (\pi/a,\pi/a)$ and $q_{2}~\sim~0.11~\cdot(\pi/a,\pi/a)$~(Figure~\ref{f3}d) based on the FS map collected for the energy of 70~eV. The obtained value of the q$_{1}$  vector is in good agreement with Ref.~\cite{24kondo_unexpected_2010} for BaFe$_{2}$As$_{2}$ and Ref.~\cite{23wang_symmetry-broken_2013} for CaFe$_{2}$As$_{2}$, being $0.33~\cdot~(\pi/a,\pi/a$). However, in Ref.~\cite{23wang_symmetry-broken_2013} there is also the value determined for cobalt doped CaFe$_{2}$As$_{2}$, equal to $0.24~\cdot~(\pi/a,\pi/a)$, which is significantly different from the obtained value. The same article suggests that this value remains constant in the whole momentum space. However, our results, as well as the map as a function of photon energy in Ref.~\cite{24kondo_unexpected_2010}, suggest FS variation with k$_{z}$, which explains the possible differences of the nesting vector between the cited ones and our data. Band topography along $\bar{\Gamma} - \bar{\text{X}}$ appears to be similar for CaFe$_{2}$As$_{2}$ (Figure~\ref{f3}e) and CaFe$_{1.93}$Co$_{0.07}$As$_{2}$ (Figure~\ref{f3}f) with the bands better resolved for the doped system. We observe three hole pockets around the $\bar{\Gamma}$ point; $\alpha$ - innermost, $\beta$ and $\gamma$ - external. The electron pocket around $\bar{\text{X}}$ ($\delta$) has its image ($\delta$') around $\bar{\Gamma}$. Such band structure corresponds to SDW state in BaFe$_{2}$As$_{2}$ system.~\cite{PhysRevB.99.035118}

Let us discuss in more detail the expected differences between the band structures of the tested compounds caused by doping. It is a matter of debate, whether transition metal doping in Fe pnictides and chalcogenides acts as carrier doping \cite{PhysRevB.83.094522, PhysRevLett.110.107007,26thirupathaiah_effect_2016, 27vilmercati_nonrigid_2016, 28liu_evidence_2010, 29liu_importance_2011} or it plays another role like contribution to scattering or increasing disorder \cite{PhysRevLett.105.157004, PhysRevLett.108.207003}. The rigid band-shifting scenario assumes a constant shift of the bands under the influence of doping, which does not result in the deformation of the band structure. In other words, doping in this simplification would only result in a shift of the Fermi energy~\cite{26thirupathaiah_effect_2016, 27vilmercati_nonrigid_2016, 28liu_evidence_2010, 29liu_importance_2011}. In Figure~\ref{f4} we present a detailed analysis of the band dispersion across the $\bar\Gamma$--$\bar\Gamma$ direction along $k_{y}$ as defined in Figure~\ref{f1}b. The first column shows the measured band dispersion collected for the undoped sample, the second shows their 2D curvature calculated using the method described in Ref.~\cite{30zhang_precise_2011}, and then analogously for the doped sample. The innermost hole-like band $\alpha$, which has higher intensity for LV polarization, is often located below $E_{F}$ with the lowest position at photon energy h$\nu$=33 eV, which confirms its dispersion along $k_z$. $\beta$ (middle) and $\gamma$ (outer) hole pockets can be visible with both polarizations but they are not always well resolved. Dimensionality of $\beta$ and $\gamma$ can be rather studied in Figure~\ref{f2} from which it is clear that $\beta$ is quite 2-dimensional, whereas $\gamma$ band is more 3-dimensional.

To determine dispersions Gaussian function was fitted to both raw data MDCs and the profiles from 2D curvature corresponding to MDCs, both shown in Figure~\ref{f4}. There is a very good agreement between the dispersions obtained from both methods of fitting and the resulting curves are typically superimposed on each other. The fit results are shown in Figure~\ref{f4} (last column). It is clearly visible that for all photon energies and both polarizations a noticeable band shift is absent within the measurement resolution limits. However, based on our previous work for Fe$_{1-x}$Co$_x$Te$_{1-y}$Se$_{y}$~\cite{31rosmus_effect_2019}, the number of electrons added to the system with 3.5\% cobalt doping per iron atom corresponds to a band shift of about 10~meV. If we assume the energy shifts of similar order in CaFe$_{2-x}$Co$_{x}$As$_{2}$ the results would be at the limit of energy resolution.

The measurements collected for the sample, which was highly doped with cobalt (CaFe$_{1.85}$Co$_{0.15}$As$_{2}$), in the tetragonal paramagnetic phase are shown in Figure~\ref{f5}. The mapping of the Fermi surface was carried out for the three different photon energies (33~eV, 50~eV, 70~eV) using horizontal and vertical polarization. The mapped  FSs are presented in Figure~\ref{f5}(a-f). They consist of concentric cylindrical $\beta2$ and $\gamma2$ hole pockets around the $\bar\Gamma$ point and electron pockets $\delta2$ and $\epsilon2$ around the $\bar{\text{X}}$ point. This structure is significantly different from that observed for the parent compound and the sample with low cobalt content (CaFe$_{1.93}$Co$_{0.07}$As$_{2}$). Therefore, our data reveal that such FS is related to the tetragonal phase of the CaFe$_{2-x}$Co$_{x}$As$_{2}$ system (Figure~\ref{f1}) in which superconductivity appears. It is compatible with FS found in earlier reports for high temperature tetragonal phase of CaFe$_{2}$As$_{2}$\cite{21liu_three-_2009, PhysRevB97054505, 17ali_emergent_2017}. Maps of the Fermi surface for different locations in the k$_{z}$ direction showed the changes of both $\beta2$ and $\gamma2$ bands around the $\Gamma$ point (Figure~\ref{f5} (a-f)). 

The band structure along the $\bar\Gamma$--$\bar{\text{X}}$ direction (Figure~\ref{f5}g) is shown with added spectra of LH and LV polarizations. High intensity $\beta2$ pocket is well visible with possible contribution from the inner $\alpha2$ band and the external $\gamma2$ pocket is also found. In Figure~\ref{f5} (h) two electron pockets around the $\bar{\text{X}}$ point; $\delta2$ and $\epsilon2$, are well resolved. The band names are altered as they concern the tetragonal phase. If we compare the band structure of highly doped systems ($x = 0.15$) with that for weakly doped ($x = 0.07$) and undoped compounds, it appears that for $x = 0.15$ the image of the electron pocket around $\Gamma$ named $\delta'$ (Figure~\ref{f3}) is absent, as this is present only in SDW phase. The spectra recorded along the $\bar\Gamma$--$\bar\Gamma$ direction (along $k_{y}$ as defined in Figure~\ref{f1}b) are presented in Figure~\ref{f5}i-n. They reveal the dispersion of $\beta2$ band along $k_z$, which is better visible with LH polarization. This can be noticed as the energy of the maximum of this band is displaced from above the Fermi energy  (e.g. Figure~\ref{f5}(k)) to the situation in which it touches E$_{F}$ (e.g. Figure~\ref{f5}(j)). This also confirms the existence of weak dispersions along $k_z$ in the quasi two-dimensional nature of the band structure. $\alpha2$ pocket is more difficult to resolve but it contributes to the spectra obtained with LV polarization. The presented band structure  is typical  of 122 family in superconducting tetragonal phase~\cite{32kordyuk_iron-based_2012,33liu_electronic_2015}. 

In the case of the parent compound and the lightly doped sample, careful analysis of the signal creating two bright spots around the $\bar{\text{X}}$ point (exact position is marked by the pink dashed line in Figure~\ref{f3}(a) and (c)) revealed the existence of a Dirac-like dispersion. The intensity plots of the ARPES spectra collected along that direction and the corresponding 2D curvatures are shown in Figure~\ref{f6}. The apex of the cone is located about 30 meV beneath the E$_{F}$, making it much deeper than reported for  BaFe$_{2}$As$_{2}$, where the apex of the cone lays above E$_{F}$~\cite{5richard_observation_2010}. The spectra recorded for the neighboring values of $k_{x}$ confirm a conical shape of the band structure at this place. The observation of these structures suggested that CaFe$_{2}$As$_{2}$ could also be classified as the nodal SDW system~\cite{13ran_nodal_2009, 34sugai_nodal_2011}. However, we do not have conclusive results concerning the existence of the nodes in a gap function. It is worth noting that the cones were observed for both samples from SDW region of the phase diagram, including the superconducting one. This may be the result of the coexistence of the superconducting phase and the Dirac-like dispersion. The possible explanation is the heterogeneity of the samples; a part of the volume can be in the SDW phase and this can be the source of the signal with cones; another part of the sample is in a superconducting phase which is visible in transport measurements. A detailed analysis of the spectra recorded in the region of the $\bar{\text{X}}$ point was also performed for highly doped CaFe$_{1.85}$Co$_{0.15}$As$_{2}$ and no dispersion, which could be related to Dirac cone, was found. This is consistent with the assumption that the occurrence of topological states in this system is related to the nature of SDW~\cite{13ran_nodal_2009, 5richard_observation_2010}, as the SDW does not appear in highly doped sample.

 \section{Summary}
\label{sec.sum}
To conclude, we have investigated the electronic structure of pure CaFe$_{2}$As$_{2}$ orthorhombic SDW system, cobalt-doped orthorhombic SDW CaFe$_{1.93}$Co$_{0.07}$As$_{2}$ located at the onset of superconductivity and higly doped tetragonal CaFe$_{1.85}$Co$_{0.15}$As$_{2}$ superconductor. We have observed two types of electron structures depending on the doping level. For SDW phase, a characteristic petal-like structure in FS related to SDW is found, while the structure in the paramagnetic phase consists of concentric hole bands around the $\bar{\Gamma}$ point and electron bands around the $\bar{\text{X}}$ point. In the band structures of all the investigated systems three hole bands around the $\bar{\Gamma}$ point are found and the innermost band is often below $E_F$. In the band structure of undoped and low doped SDW systems electron pocket at $\bar{\text{X}}$ and its replica at $\bar{\Gamma}$ are visible. All the studied bands are quasi 2-dimensional with observed weak dispersions along $k_z$. We did not find a significant change in the band structure under the influence of cobalt doping within the orthorhombic SDW phase. For CaFe$_{2}$As$_{2}$ and CaFe$_{1.93}$Co$_{0.07}$As$_{2}$ systems, in the SDW phase, the existence of Dirac cones was revealed. In contrast to BaFe$_{2}$As$_{2}$, the Dirac cones are located 30 meV below $E_F$. 

\section{Acknowledgments}
This publication was partially developed under the provision of the Polish Ministry and Higher Education project ``Support for research and development with the use of research infra-structure of the National Synchrotron Radiation Centre SOLARIS'' under contract nr 1/SOL/2021/2. 
We acknowledge SOLARIS Centre for the access to the URANOS beamline, where the measurements were performed.

\bibliography{CaFeAs}%
\end{document}